\documentclass[journal,twocolumn]{IEEEtranTCOM}
\usepackage{array,cite}
\usepackage{graphicx,epsfig,subfigure}
\usepackage{amsthm}
\usepackage{amsmath}
\usepackage{mdwmath}
\usepackage{array}
\usepackage{subeqnarray}
\usepackage{stfloats}
\usepackage{algorithm,algorithmic}
\usepackage{xcolor}

\makeatletter
\newcommand{\rmnum}[1]{\romannumeral #1}
\newcommand{\Rmnum}[1]{\expandafter\@slowromancap\romannumeral #1@}
\makeatother

\newtheorem{theorem}{Theorem}
\newtheorem{lemma}{Lemma}
\newtheorem{corollary}{Corollary}

\IEEEoverridecommandlockouts

\begin{document}

\title{Virtual-MIMO-Boosted Information Propagation on Highways}

\author{\IEEEauthorblockN{Zhaoyang~Zhang, \textit{Member, IEEE}, Hui~Wu, Huazi~Zhang, \textit{Member, IEEE}, \\Huaiyu Dai, \textit{Senior Member, IEEE}, and Nei Kato, \textit{Fellow, IEEE}}
\thanks{
This work was supported in part by National Key Basic Research Program of China (No. 2012CB316104), National Hi-Tech R\&D Program (No. 2014AA01A702), Zhejiang Provincial Natural Science Foundation of China (No. LR12F01002), National Natural Science Foundation of China (Nos. 61371094, 61401388) and the China Postdoctoral Science Foundation Funded Project (No. 2014M551736).

Z. Zhang, H. Wu, and H. Zhang (E-mails: {\tt \{ning\_ming, 3090102503, hzhang17\}@zju.edu.cn}) are with the Department of Information Science and Electronic Engineering, Zhejiang University, China. H. Dai (Email: {\tt hdai@ncsu.edu}) is with the ECE Department, NC State University, USA. N. Kato (Email: {\tt kato@it.ecei.tohoku.ac.jp}) is with the Graduate School of Information Sciences, Tohoku University, Japan.}
}

\maketitle

\begin{abstract}
In vehicular communications, traffic-related information should be spread over the network as quickly as possible to maintain a safer transportation system. This motivates us to develop more efficient information propagation schemes. In this paper, we propose a novel virtual-MIMO-enabled information dissemination scheme, in which the vehicles opportunistically form virtual antenna arrays to boost the transmission range and therefore accelerate information propagation along the highway. We model the information propagation process as a renewal reward process and investigate in detail the \emph{Information Propagation Speed} (IPS) of the proposed scheme. The corresponding closed-form IPS is derived, which shows that the IPS increases cubically with the vehicle density but will ultimately converge to a constant upper bound. Moreover, increased mobility also facilitates the information spreading by offering more communication opportunities. However, the limited network density essentially determines the bottleneck in information spreading. Extensive simulations are carried out to verify our analysis. We also show that the proposed scheme exhibits a significant IPS gain over its conventional counterpart.
\end{abstract}


\section{Introduction}
\subsection{Motivation}
Vehicular ad hoc networks (VANET) is undergoing extensive study in recent years \cite{model1,model2,model3,add1,add2}. Maintaining a safer transportation system is of top priority. One safety measure is to enable inter-vehicle communication, especially in highways without any fixed road-side infrastructure. Information on traffic-related events, including accident, traffic jam, closed road, etc, needs to be relayed to nearby vehicles in a multi-hop fashion. For instance, if an accident occurs on a highway and induces temporary congestion, it is vital to ensure that all vehicles in this region be informed as quickly as possible, so that some of them can make early detours before getting trapped. Therefore, we are interested in a new performance metric that describes how fast a message can propagate along the route. This new metric, \emph{Information Propagation Speed} (IPS) \cite{varyv2}, defined as the distance the information spreads within unit time, has drawn increasing attention recently.

\subsection{Related Work}
The allocation of a 75 MHz licensed band at 5.9 GHz for Dedicated Short Range Communications (DSRC) \cite{DSRC}, which was later standardized
by IEEE 802.11p in vehicular environments \cite{IEEE802}, has renewed interest from both industry and academia in VANET, which usually supports high and yet regular and predictable mobility, due to road and traffic constraints (e.g. vehicles traveling on highways). In this context, early works \cite{model1,model2,model3,model4,model5} focused on modeling various aspects of VANET. In \cite{model4}, the authors studied vehicle traces and concluded that vehicles are very close to being exponentially distributed in highways. Further, measurements in \cite{model5} showed that vehicles traveling in different lanes (e.g. bus lane or heavy truck lane) have different speed distributions.

Information spreading in mobile networks has become a research hot spot recently. On the one hand, some interesting scaling properties have emerged, concerning mobile ad hoc networks (MANET). For instance, \cite{HZ1,HZ2} analyzed the effect of mobility on information spreading in large-scale mobile networks. On the other hand, in VANET, information spreading also has wide applications, among which the emergency message dissemination is of major importance. On the other hand, in VANET, information spreading also has wide applications, among which the emergency message dissemination is of major importance. Different applications are likely to have different requirements for the corresponding information flow. For instance, safety applications are expected to be low data-rate, confined to a small neighbourhood, typically of the order of several hundred meters, but with strict latency constraints (of the order of few milliseconds). This motivates works dedicated to reliable emergency message exchange \cite{congestioncontrol},\cite{tlc}. Comparatively, other traffic related services such as traffic monitoring etc., tends to have more relaxed latency constraints, and require communication within neighborhood spanning several or tens of kilometers. In this case, continuous and complete connectivity may be unavailable due to the frequent-changing network topology, thus a store-and-forward scheme is usually adopted to allow messages to be stored in the memory of a node and forwarded when possible and necessary. Therefore, in such delay tolerant vehicular networks, many researchers are more interested in the evaluation of IPS.
Agarwal et al. \cite{upblb1,upblb2} first obtained upper and lower bounds for the IPS in an 1-D VANET, where vehicles are Poissonly distributed and move at the same, constant speed in opposing directions. Subsequently, Baccelli et al. derived more fine-grained results on the limit \cite{exactphasetransi}, under the same setting of \cite{upblb2}. The authors complemented their work by taking into account radio communication range variations at the MAC layer, and characterized conditions for the phase transition \cite{varyradiorange}.

Other than traffic density, the impact of vehicle speed on IPS has also been extensively investigated. In \cite{wu}, Wu et al. studied the IPS assuming uniformly distributed, but time-invariant vehicle speeds, and obtained analytical results in extremely sparse and extremely dense cases. \cite{varyv1} showed that the time-variation of vehicle speed is also closely related to IPS. When extended to a multi-lane scenario, \cite{varyv2} obtained similar results. Note that in all the above works, each vehicle plays the role of an individual relay, which communicates with only one other vehicle at a time, and no inter-vehicle cooperation has been considered.

Notice that most existing literature mentioned above focus on the evaluation of IPS under various analytical models \cite{upblb1}-\cite{varyv2}, where the routing schemes are usually based on the simplest point-to-point transmission, i.e., each vehicle only communicates with one other vehicle within its own transmission range \cite{wu}-\cite{varyv2}. Particularly, when considering a typical bidirectional setting, \cite{upblb1}-\cite{exactphasetransi} adopted a routing scheme which uses reverse traffic opportunistically to maintain a transient connectivity of the whole network. To be more specific, upstream messages are forwarded by vehicles traveling in the same direction, until a partition in the upstream traffic is encountered. Then the downstream traffic to utilized to bridge traffic clusters (i.e., maximal sequence of vehicles such that two consecutive cars are within communication range) and forward the message to the next upstream cluster. Note that in all above works, each vehicle plays the role of an individual relay, and no inter-vehicle cooperation has been considered. Innovatively, our previous work, based on a similar bi-directional model, proposed a cluster-based cooperative message forwarding scheme, which involves multiple transmitters and multiple receivers for each cluster-to-cluster transmission, and dramatically enhances the IPS by exploiting the power and diversity gain offered by virtual MIMO arrays. However, \cite{faster} assumes identical vehicle densities and instantaneous intra-cluster transmission, which are somewhat simplistic. Therefore, we are motivated to build a model with more realistic assumptions and develop more efficient information propagation schemes.
\subsection{Summary of Contributions}
In this paper, we study the information propagation in a bidirectional highway scenario. We propose a virtual-MIMO-enabled information dissemination scheme and analyze the corresponding IPS. We show that the proposed scheme yields faster information propagation than conventional ones. Our main contributions are summarized below.
\begin{enumerate}
\item{Compared with the existing highway information propagation schemes, the novelty of ours lies in the following aspects:
  (i) Instead of the widely-adopted unicast mechanism, we alternatively adopt an ACK-free epidemic broadcast protocol, which will not only reduce communication overheads but also avoid unpredictable delays (particularly in some error-prone conditions).
  (ii) We allow signals to travel between traffic lanes, so that more communication opportunities can be exploited to accelerate the dissemination process.
  (iii) To further improve information propagation performance, we allow an individual vehicle to aggressively combine as many signals as it can detect, which can increase the probability of successful decoding.}
\item{We model the information propagation process as a renewal reward process and evaluate in detail the IPS of the proposed scheme. Closed-form results are obtained, and further analysis reveals that the IPS increases cubically with the vehicle density but will ultimately converge to a fixed upper bound. Meanwhile, mobility can also facilitate information propagation by offering more communication opportunities. Most importantly, it can be observed that our scheme significantly outperforms others, exhibiting a remarkable IPS gain.}
\end{enumerate}

The rest of the paper is organized as follows - In Section \ref{systemmodel}, we present the system model and the main idea of our proposed virtual-MIMO-enabled propagation scheme. In Section \ref{IPSvmimosec}, we derive the closed-form expression of IPS for the proposed scheme and its asymptotic properties will be addressed in Section \ref{furtheranalysis}. Simulations results are given in Section \ref{simulation}. The paper is finally concluded in Section \ref{conclusion}.
\section{System Model and Main Idea}\label{systemmodel}

\subsection{Network Model}

\begin{figure}[!ht]
\center
\subfigure[At the beginning of the $i$th slot, the current head is \textbf{b}.]{\includegraphics[angle=0,scale=0.35]{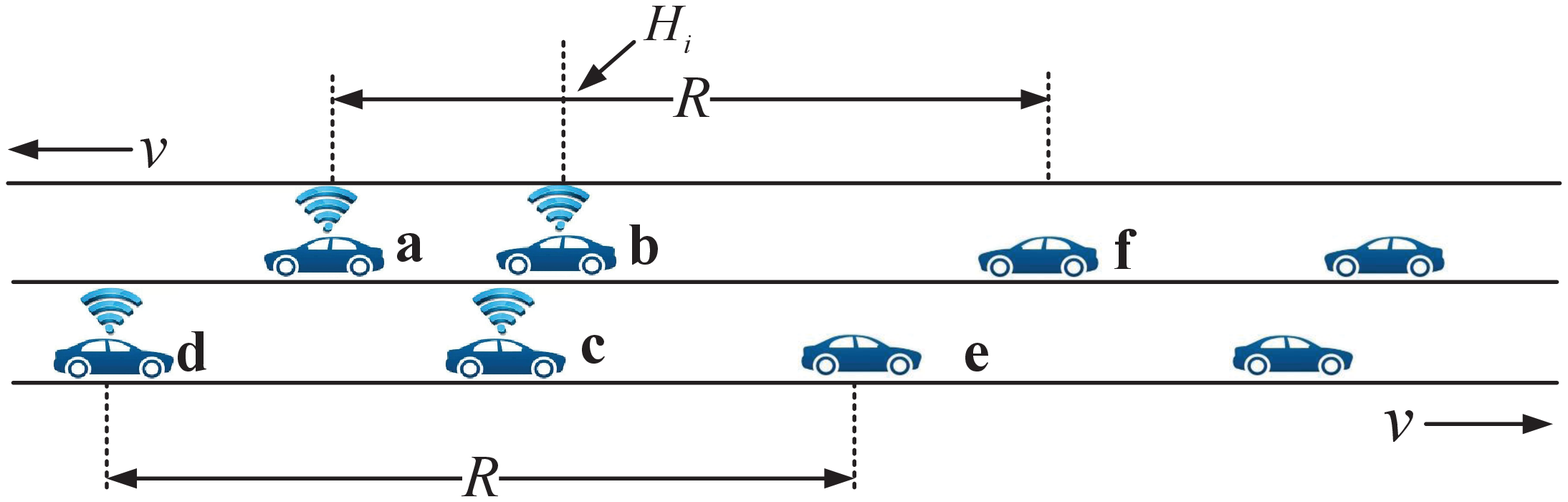}\label{sloti}}
\subfigure[During the $i$th slot, \textbf{e} and \textbf{f} successfully receive the message.]{\includegraphics[angle=0,scale=0.35]{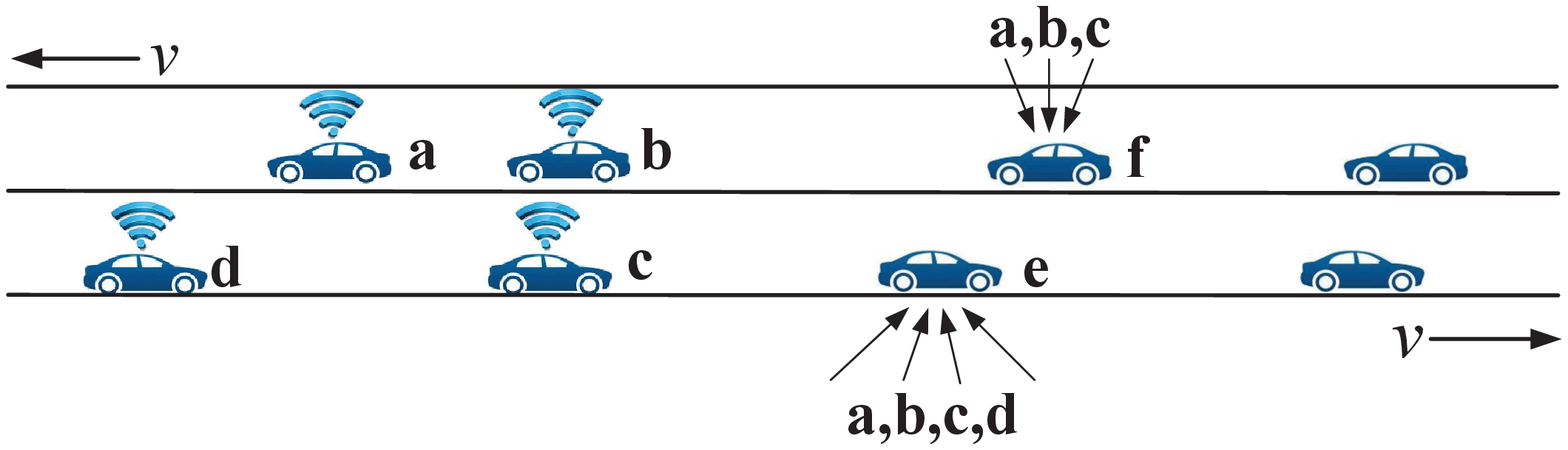}\label{slotitoi1}}
\subfigure[At the beginning of the $(i{+}1)$th slot, \textbf{f} becomes the new head.]{\includegraphics[angle=0,scale=0.35]{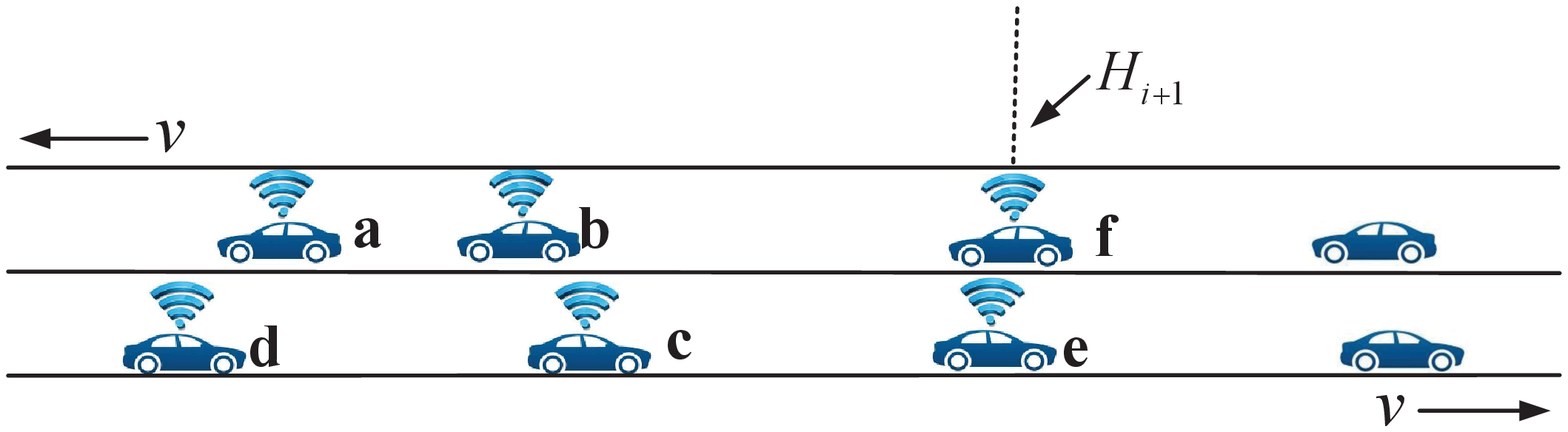}\label{sloti1}}
\caption{Illustration of the information propagation process over a highway vehicular network.}
\label{Head}
\end{figure}

As depicted in Fig. \ref{Head}, we consider a highway scenario where vehicles in two lanes move in opposite directions at a constant speed $v$. We assume that both eastbound and westbound vehicles are Poisson distributed, which is shown to be a reasonable approximation on non-congested highways \cite{poisson}. Without loss of generality, we focus on the problem of unidirectional dissemination of single-piece small-sized information (\emph{the beacon}) which deserves being exclusively disseminated in the highway network, e.g., a piece of emergency message, issued by a westbound vehicle and to be propagated eastward. In the following, the westbound lane is referred to as the \emph{reverse lane} where vehicles travel in the opposite direction of the beacon, while the eastbound lane is referred to as the \emph{forward lane}. Let $\lambda_r$ and $\lambda_f$ denote the Poisson densities of the two lanes respectively.

It is noteworthy that synchronization is crucial to the studied information propagation scheme. In this work, by assuming the vehicles are equipped with some global timing facilities like GPS (Global Positioning System) as widely deployed in many systems nowadays, it is easy to achieve network-wide time synchronization. In case there is no global timing facility available, some practical algorithms based on time-stamp exchange and update \cite{sync1} can also be used to achieve this goal. In addition, more precise frame-level and symbol-level synchronization can be accomplished by employing algorithms developed for distributed MIMO systems based on the specially designed preambles as elaborated in \cite{sync2,sync3}, in which the preambles are designated to be capable of distinguishing the starting point of each node's frame signal that experiences uncertain propagation delay and channel realization. For those newly-joined vehicles, we can allow a a warm-up period for them to learn the sync pattern incumbent transmissions, and only those well-synchronized cars (which pass a sync test) can join the communication. Therefore, hereafter, we assume synchronization is achieved.

\begin{figure}[!ht]
\centering
\includegraphics[angle=0,scale=0.35]{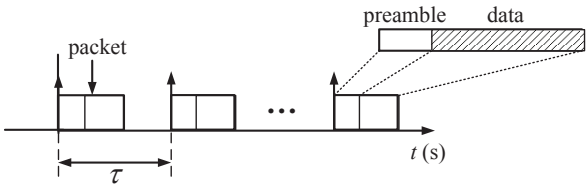}
\caption{Time is equally divided into slots of $\tau$ seconds.}
\label{timeslots}
\end{figure}

We examine the dissemination problem over discrete time steps, as shown in Fig. \ref{timeslots}. We assume that a vehicle with the emergency message, referred to as the \emph{informed vehicle}, will repeatedly broadcast the message at each time slot of duration $\tau$ and with equal fixed power $P_t$. To facilitate our further analysis, it is further assumed that the message packet of each vehicle contains a specific preamble segment used for identification, synchronization and channel estimation, and all transmission links work in the half-duplex mode. Additionally, it is worth mentioning that the time slot duration is set sufficiently long enough to not only cover the transmission delay and propagation delay of the short-sized beacon/packet (usually a few bits is enough to indicate critical information such as time and location), but also enable essential signal processing such as synchronization, interference elimination, combining and decoding, etc.

\subsection{Virtual-MIMO-Enabled Information Propagation}\label{vmimo}
Before elaborating the proposed scheme, let us first look into the transmission process of a single pair of transmitter and receiver. Denote by $x$ the transmitted beacon, then the signal received by an uninformed vehicle from one of its neighboring informed vehicles is
\begin{equation}
y=hx+w,
\end{equation}
where $h$ is the channel coefficient between the vehicles, and $w$ is the received additive white gaussian noise (AWGN) with power $N_0$. Considering the open space nature of most vehicular environments, we adopt a free space model to characterize all the inter-vehicle channels, i.e., $|h|^2\propto\frac{1}{d^2}$, where $d$ is the transmitter-receiver distance. Hence, the received signal-to-noise ratio (SNR) is
\begin{equation}
\label{channelmodel}
\gamma=\frac{\alpha P_t}{N_0 d^2},
\end{equation}
where $\alpha$ is a constant associated with antenna characteristics.


Define two fixed SNR thresholds: the \emph{decoding threshold} $\gamma_{dec}$ and the \emph{detecting threshold} $\gamma_{det}$ ($\gamma_{det} \le \gamma_{dec}$), respectively, which correspond to two different ranges, namely, the \emph{transmission range} $r$ and the \emph{detection range} $R$ ($R>r$), respectively, given the transmit power and the received noise. The transmission range $r$ means the maximum distance between two communicating vehicles when the beacon can be successfully decoded (or $\gamma\geq\gamma_{dec}$), while the detection range $R$ means the maximum distance for the receiving vehicle to successfully detect the signal from the transmitting vehicle (or $\gamma\geq\gamma_{det}$) (see Fig.\ref{slotitoi1}). As a matter of fact, vehicles that can make meaningful contacts and thus essentially contribute to the information dissemination lie within a distance of no farther than $R$ from each other. The reason is that, an uninformed vehicle beyond a distance $R$ from the informed vehicles has no chance of detecting the beacon, let alone decoding it, while an informed vehicle beyond $R$ from the uninformed vehicles has no chance of being detected at all.

Simply put, the main idea of our proposed virtual-MIMO-enabled information propagation is to allow individual vehicles to aggressively combine as many signals as they can detect within their own detection ranges, so as to achieve successful decoding as early as possible. The process is illustrated in Fig. \ref{Head}. At the beginning of a certain time slot, all the informed vehicles (such as $\textbf{a}$, $\textbf{b}$, $\textbf{c}$, $\textbf{d}$ in Fig.\ref{sloti}) independently and simultaneously broadcast the beacon. The nearby uninformed vehicles within the detection range (such as $\textbf{e}$ and $\textbf{f}$ in Fig.\ref{sloti}) are able to overhear some of these signals. Take \textbf{f} for example. As illustrated in Fig.\ref{slotitoi1}, it can hear three signals \{$h_ix$\} simultaneously sent from \textbf{a},\textbf{b} and \textbf{c}, respectively, where \{$h_i$\} are the respective channel coefficients which can be viewed as independent as long as the distances between vehicles are much longer than the wavelength. That means, the signal received by \textbf{f} is
\begin{equation}
y_f = \sum \limits_{i\in\{a,b,c\}} h_i x + w.
\end{equation}
By using the dedicated preamble of each branch signal, \textbf{f} estimates the channel coefficients $\{h_i\}$. Then in a way similar to the traditional maximal-ratio combining (MRC) \cite{wc}, it multiplies the received signal with weights $\{\frac{h^*_i}{\sum \limits_{i} |h_i|^2}\}$ and then linearly combines them together, which yields
\begin{equation}
\bar{y}=\sum \limits_{i} \frac{h^*_i}{\sum \limits_{i} |h_i|^2} y_f.
\end{equation}
The resultant average SNR of $\bar{y}$ is just $\gamma = \frac{\alpha P_t \sum\limits_{i}{|h_{i}|^2}}{N_0}$. If it exceeds $\gamma_{dec}$, \textbf{f} will be informed. Note that following a similar procedure, other uninformed vehicles (such as \textbf{e}) may also yield successful receptions at the same time, thus making it a virtual MIMO scenario.

Define \emph{head} as the rightmost informed node at each time slot. Fig. \ref{Head} shows an example of a single-slot propagation: let $H_i$ be the head at the beginning of the $i$th slot (i.e., \textbf{b}). During the slot, \textbf{e} and \textbf{f} are able to decode the message packet through the above virtual MIMO enhanced approach. Thus, when the next time slot begins, \textbf{f} becomes the new head, denoted by $H_{i+1}$. Therefore, the information propagation process can be captured by the evolvement of the head coordinate, which will be discussed in detail in Section \ref{IPSvmimosec}.

Apparently, the scheme can disseminate the information faster, by allowing vehicles beyond $r$ to opportunistically decode the message as well.
Besides, it is also practical for one-dimensional spatially-distributed vehicular networks, without any complex signal processing techniques (e.g. space-time block code (STBC)) in most virtual-MIMO-based systems. Instead, vehicles work simply and independently, either broadcasting or receiving and combining signals, yet still benefit from virtual antenna arrays. Moreover, to the best of our knowledge, it is the first time that the both lanes are given the same priority to be informed of the emergency, concerning a bidirectional highway network, which better exploits the dynamics of the network topology to improve the system performance.

\section{Information Propagation Speed Evaluation}
\label{IPSvmimosec}
We now analyze the IPS, the key performance metric of our proposed scheme, by establishing a tractable analytical model for the information dissemination process.
%
%
%

\subsection{Analytical Model}

As mentioned above, the propagation progress can be essentially captured by the evolving coordinate of the head. This key observation leads to an effective way to evaluate the IPS. To facilitate the analysis, vehicles in the reverse lane are viewed as motionless nodes, while those in the forward lane are moving at a speed of $2v$. Our main goal is to quantify the relative IPS with respect to the westbound vehicles.

\begin{figure}[!ht]
\center
\subfigure[The beacon keeps propagating forward.]{\includegraphics[angle=0,scale=0.285]{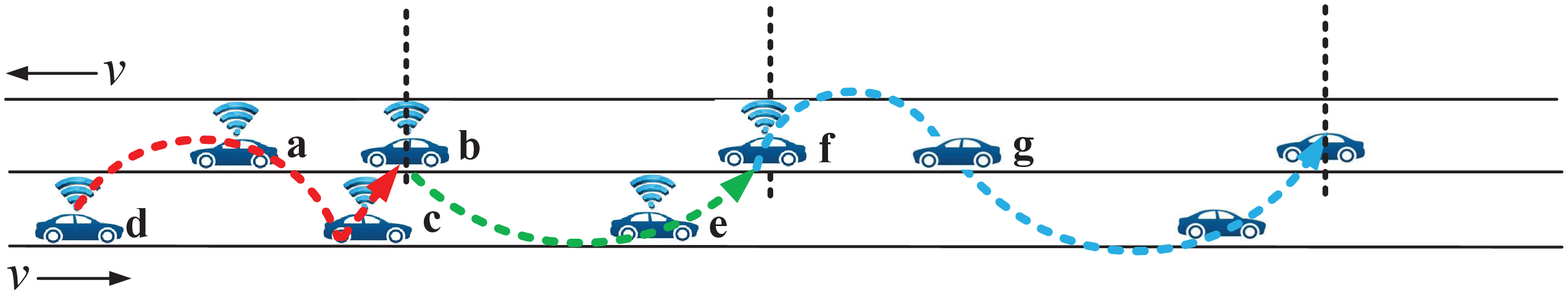}\label{hoppingcase}}
\subfigure[The beacon is blocked in the forward lane.]{\includegraphics[angle=0,scale=0.285]{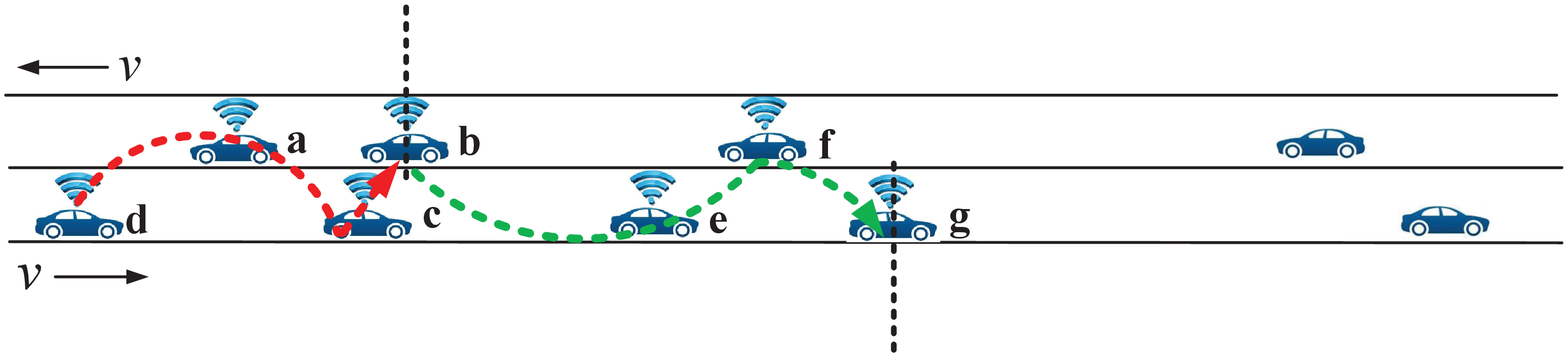}\label{carrycase}}
\subfigure[The beacon is blocked in the reverse lane.]{\includegraphics[angle=0,scale=0.285]{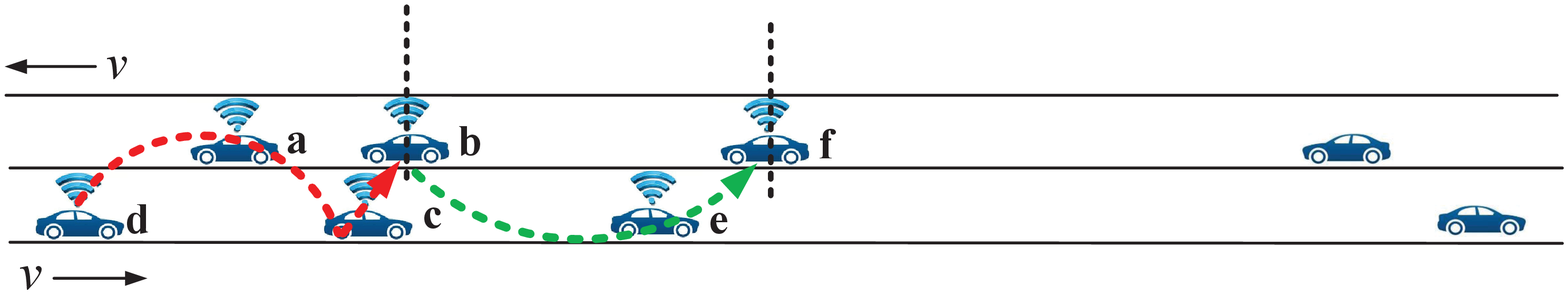}\label{stopcase}}
\caption{The information propagation pattern.}
\label{propagationpattern}
\end{figure}

\begin{figure}[!ht]
\centering
\includegraphics[angle=0,scale=0.25]{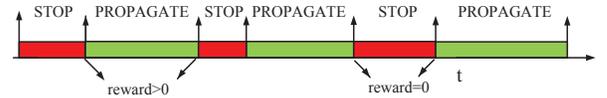}
\caption{Renewal reward process.}
\label{renewal}
\end{figure}

As in Fig. \ref{hoppingcase}, the beacon keeps propagating forward in a new time slot, if the gap between the head and the nearby uninformed vehicles can be bridged by our virtual MIMO approach. The rightmost newly informed vehicle, as indicated by the upright dotted line, thus becomes the new head of the current time slot. Otherwise, if a large gap is encountered, the beacon will be blocked and thus have to wait in the current head, either in the forward lane (see Fig. \ref{carrycase}) or the reverse lane (see Fig. \ref{stopcase}), until connectivity is restored. Concerning the relative IPS with respect to the reverse traffic, we can see that only when the beacon is blocked in the reverse lane shall it be considered a halt while in other cases it will remain propagating. Therefore, we can model the whole propagation process as a renewal reward process with two alternating states, i.e., the PROPAGATE state and the STOP state, as illustrated in Fig. \ref{renewal}. Here the reward of each cycle can be interpreted as the relative distance the beacon traverses across the heads.

\begin{figure}[!ht]
\centering
\includegraphics[angle=0,scale=0.3]{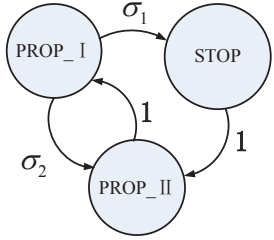}
\caption{The whole propagation can be modeled by a continuous Markov chain.}
\label{whole markov}
\end{figure}

Due to the linear mobility model and the Poisson traffic model, the network can be viewed as stationary. Thus, a continuous Markov chain with each state lasting for a certain number of time slots, as shown in Fig. \ref{whole markov}, can be adopted to characterize the whole information propagation process. The state machine of the chain can be determined from the following observations. First, the PROPAGATE state in Fig. \ref{renewal} can be viewed as a combination of two sub-states, denoted as PROP\_\Rmnum{1} and PROP\_\Rmnum{2}, respectively, with the former indicating that the beacon keeps propagating along and/or across the lanes, and the latter indicating that it is blocked in the forward lane but still maintains a positive propagating speed of 2$v$. Second, once the system leaves the PROP\_\Rmnum{1} state due to a large gap, it will transit either to the PROP\_\Rmnum{2} state with some probability $\sigma_1$ or to the STOP state with some probability $\sigma_2$, depending on whether the beacon is blocked in the forward lane or the reverse lane, respectively. Third, the PROP\_\Rmnum{2} state will always evolve to the PROP\_\Rmnum{1} state (with probability 1), since the moving head will eventually approach some reverse lane vehicle ahead and pass on the beacon to it. Similarly, the STOP state will always evolve to the PROP\_\Rmnum{2} state, since a reverse lane head will eventually be overtaken by some moving vehicle in the forward lane, which then serves as the new head.

\begin{figure}[!ht]
\centering
\includegraphics[angle=0,scale=0.25]{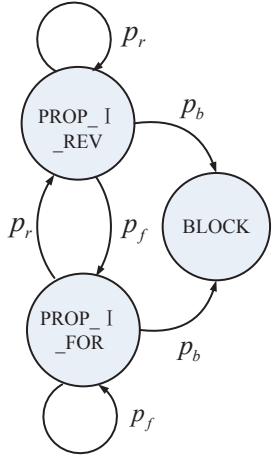}
\caption{The PROP\_I state is modeled by a discrete Markov chain.}
\label{hoppingmarkov}
\end{figure}

Moreover, in the PROP\_\Rmnum{1} state, the head position evolves slot by slot, depending on the distribution of the current informed vehicles and the nearby uninformed ones within the detection ranges. Therefore, this state can be further characterized by another discrete Markov chain as depicted in Fig. \ref{hoppingmarkov}, where its two substates, PROP\_\Rmnum{1}\_REV and PROP\_\Rmnum{1}\_FOR, denote that after each time slot the new head lies in the reverse lane and the forward lane, respectively, with $p_f$ and $p_r$ denoting the corresponding transition probabilities. BLOCK denotes an absorbing state in which the beacon is blocked by a gap and thus the system temporally leaves the current PROP\_\Rmnum{1} state. Since the nearby uninformed vehicles are independently Poisson distributed in the two lanes given the locations of the current informed ones, the probability of the next head lying in either lane is proportional to the corresponding vehicle density. Hence, there is $p_r+p_f = 1-p_b$ and $p_r:p_f =\lambda_r : \lambda_f$, which yields
\begin{equation}\label{pf}
p_f=(1-p_b)\cdot \frac{\lambda_f}{\lambda_r+\lambda_f},
\end{equation}
\begin{equation}\label{pr}
p_r=(1-p_b)\cdot \frac{\lambda_r}{\lambda_r+\lambda_f},
\end{equation}
where $p_b$, the probability of being blocked, is given by the following lemma:
\begin{lemma}
\label{pb}
The probability $p_b$ that the beacon is blocked, i.e., it cannot propagate any further, is:
\begin{equation}
\label{pbequa}
p_b = e^{-\lambda r} \Phi \left(\frac{\frac{1}{r^2}-\mu}{\sigma}\right) = \Theta \left( {{e^{ -\beta \lambda }}} \right),
\end{equation}
where $\Phi\left(\cdot\right)$ denotes the cumulative distribution function (CDF) of the standard normal distribution, $\lambda=\lambda_r+\lambda_f$, $\mu=\lambda \left(\frac{1}{r}-\frac{1}{R}\right)$, $\sigma=\sqrt{\frac{1}{3} \lambda \left(\frac{1}{r^3}-\frac{1}{R^3}\right)}$, and $\beta>r$ is some positive constant.
\end{lemma}

\begin{IEEEproof}
See Appendix \ref{proof_lemma1}.
\end{IEEEproof}


Now, let us return to the continuous Markov chain in Fig. \ref{whole markov}, which further implies that the transition probability $\sigma_1$ ($\sigma_2$) is identical to the conditional probability that given the initial state PROP\_\Rmnum{1}\_FOR, the last state before absorption in Fig. \ref{hoppingmarkov} is PROP\_\Rmnum{1}\_REV (resp. PROP\_\Rmnum{1}\_FOR). To compute $\sigma_1$ and $\sigma_2$, we first look into $\sigma_{1m}$, the probability that after $m$ steps from the initial state PROP\_\Rmnum{1}\_FOR the system arrives the absorbing state via PROP\_\Rmnum{1}\_REV, which is
\begin{equation}
\sigma_{1m} = \left(1-p_b\right)^{m-2} p_r p_b,~m=2,3,...,
\end{equation}
where $\left(1-p_b\right)^{m-2}$ accounts for the first $(m-2)$ steps not being absorbed, while $p_r$ and $p_b$ are for the last two states PROP\_\Rmnum{1}\_REV and BLOCK, respectively. Thus we have
\begin{equation}\label{s1}
\begin{aligned}
\sigma_1  &= \sum\limits_{m = 2}^{ + \infty } \sigma_{1m} = \sum\limits_{m = 2}^{ + \infty } {{{\left( {1 - {p_b}} \right)}^{m - 2}}{p_r}{p_b}} \\
          &= {p_r} = \frac{{{\lambda _r}}}{{{\lambda _r} + {\lambda _f}}} - \frac{{{\lambda _r}}}{{{\lambda _r} + {\lambda _f}}}{p_b},
\end{aligned}
\end{equation}
\begin{equation}\label{s2}
\sigma_2 = 1-\sigma_1 =\frac{{{\lambda_f}}}{{{\lambda_r} + {\lambda_f}}} + \frac{{{\lambda_r}}}{{{\lambda _r} + {\lambda _f}}}{p_b}.
\end{equation}

\subsection{Evaluation of IPS} \label{evalips}
By the virtue of renewal reward process, the IPS can be defined as the average distance traversed within a time unit (i.e., the long-run average reward):
\begin{equation}
\label{vp}
V_\mathrm{prop\_VMIMO}=\frac{E\left[D\right]}{E\left[T\right]}=\frac{E[D_{\rm{prop}}]}{E[T_{\rm{prop}}]+E[T_{\rm{stop}}]},
\end{equation}
where $E[D_{\rm{prop}}]$ is the expected distance traversed in each cycle, and $E[T_{\rm{prop}}]$ and $E[T_{\rm{stop}}]$ are the time durations of the PROPAGATE state and the STOP state, respectively. Therefore, the remaining effort is to separately evaluate the above three expectations.

\begin{lemma}
\label{Tstop}
The expected duration of the stop state is given by:
\begin{equation}
\label{ETstop}
E\left[T_\mathrm{stop}\right]=\frac{1}{2v\lambda},
\end{equation}
where $\lambda=\lambda_r+\lambda_f$.
\end{lemma}
\begin{IEEEproof}
To ease the analysis, we only consider the worst case where the stop state will last until \textbf{e} is overtaken by the nearest informed vehicle. Thus the average distance to be covered by the bidirectional traffic is $\frac{1}{\lambda}$, i.e., the expected inter-vehicle distance. Straightforwardly, the expected duration of the stop state, i.e., the average time needed for the catch-up is given by:
\begin{equation}
\label{Tstopeq}
E\left[ T_\mathrm{stop}\right]=\frac{1}{2v\lambda}.
\end{equation}
\end{IEEEproof}

\begin{lemma}
\label{TcDc}
The expected duration and the distance traversed by the beacon of the PROP\_\Rmnum{2} state are given by:
\begin{equation}
\label{Tpause}
E\left[T_{\rm{prop\_\rmnum{2}}}\right]=\frac{1}{2v\lambda}
\end{equation}
\begin{equation}
\label{Dpause}
E\left[D_{\rm{prop\_\rmnum{2}}}\right]=\frac{1}{\lambda},
\end{equation}
where $\lambda=\lambda_r+\lambda_f$.
\end{lemma}

\begin{IEEEproof}
In the PROP\_\Rmnum{2} state, an eastbound head will store the beacon, until it approaches the nearest vehicle ahead and pass it on. Thus, the expected distance to be covered equals $E[d|d>r]-r$, yielding Eq. \eqref{Dpause}. Thus Eq. \eqref{Tpause} is obtained by $E[T_{\rm{prop\_\rmnum{2}}}]=\frac{E[D_{\rm{prop\_\rmnum{2}}}]}{2v}$.
\end{IEEEproof}

\begin{lemma}
\label{TfDf}
The expected duration and the distance traversed of the PROPAGATE state are given by:
\begin{equation}
\label{Tfeq}
E[{T_{{\rm{prop}}}}] = \left( {\frac{{{\sigma _2}}}{{{\sigma _1}}} + {\sigma _1}} \right) \times \frac{\tau}{{{p_b}}} + \frac{{{\sigma _2}}}{{{\sigma _1}}} \cdot \frac{1}{{2v\lambda}}
\end{equation}
\begin{equation}
\label{Dfeq}
E[{D_{{\rm{prop}}}}]=\left( {\frac{{{\sigma _2}}}{{{\sigma _1}}} + {\sigma _1}} \right) \times \frac{\frac{1}{{\frac{1}{{\lambda {r^2}}} + \frac{1}{R}}}}{p_b} + \frac{\sigma _2}{\sigma _1}\cdot \frac{1}{\lambda},
\end{equation}
where $\sigma _1$ and $\sigma _2$ are given by Eq. \eqref{s1} and Eq. \eqref{s2} respectively. $p_b$ is given by Eq. \eqref{pbequa}.
\end{lemma}

\begin{IEEEproof}
Apparently, for the discrete Markov chain in Fig. \ref{hoppingmarkov}, the following equation holds:
\begin{equation}
\label{absorbeq}
\left[ {\begin{array}{*{20}{c}}
{{T_r}}\\
{{T_f}}\\
{{T_b}}
\end{array}} \right] = \left[ {\begin{array}{*{20}{c}}
1\\
1\\
0
\end{array}} \right] + \Pi  \cdot \left[ {\begin{array}{*{20}{c}}
{{T_r}}\\
{{T_f}}\\
0
\end{array}} \right],
\end{equation}
where $\Pi$ is the transition matrix, and $T_r$, $T_f$, $T_b$ denote the expected absorbing time given that the initial state is PROP\_\Rmnum{1}\_REV, PROP\_\Rmnum{1}\_FOR and the absorbing state BLOCK, respectively.

By solving Eq. \eqref{absorbeq}, we have $T_r=T_f=\frac{1}{p_b}$. Note that the expected absorbing steps is equivalent to the expected number of time slots of the PROP\_\Rmnum{1} state, so we have:
\begin{equation}
\label{Thoping}
E[{T_{{\rm{prop\_\rmnum{1}}}}}] = \frac{\tau}{{{p_b}}}.
\end{equation}
Accordingly, the expected distance traversed in the PROP\_\Rmnum{1} state, is given by:
\begin{equation}
\label{dprop1}
E[{D_{{\rm{prop\_\rmnum{1}}}}}]=\frac{E\left[D_{\rm{mprop}}\right]}{p_b},
\end{equation}
where $E\left[D_{\rm{mprop}}\right]$ is the expected distance of a single-slot propagation and can be determined by Lemma \ref{Dhopping} in Appendix \ref{proof_lemma5}.

Back to the continuous Markov chain in Fig. \ref{whole markov}, we have:
\begin{equation}
\label{Tfderi}
\begin{aligned}
E[{T_{{\rm{prop}}}}] &= {\sigma _1} {\times} E[{T_{{\rm{prop\_\rmnum{1}}}}}] {+} \left( {E[{T_{{\rm{prop\_\rmnum{1}}}}}]{+} E[{T_{{\rm{prop\_\rmnum{2}}}}}]} \right)\\
 &{\times} {\sigma _1}{\sigma _2} {\times} \mathop {\lim }\limits_{n \to \infty } \sum\limits_{i = 1}^n {i{\sigma _2}^{i - 1}} \\
 &= \left( {\frac{{{\sigma _2}}}{{{\sigma _1}}} + {\sigma _1}} \right) \times E[{T_{{\rm{prop\_\rmnum{1}}}}}] + \frac{\sigma _2}{\sigma _1}E[{T_{{\rm{prop\_\rmnum{2}}}}}]
\end{aligned}
\end{equation}

\begin{equation}
\label{Dfderi}
\begin{aligned}
E[{D_{{\rm{prop}}}}] &= {\sigma _1} {\times} E[{D_{{\rm{prop\_\rmnum{1}}}}}] {+} \left( {E[{D_{{\rm{prop\_\rmnum{1}}}}}]{+} E[{D_{{\rm{prop\_\rmnum{2}}}}}]} \right)\\
 &{\times} {\sigma _1}{\sigma _2} {\times} \mathop {\lim }\limits_{n \to \infty } \sum\limits_{i = 1}^n {i{\sigma _2}^{i - 1}} \\
 &= \left( {\frac{{{\sigma _2}}}{{{\sigma _1}}} + {\sigma _1}} \right) \times E[{D_{{\rm{prop\_\rmnum{1}}}}}] + \frac{\sigma _2}{\sigma _1}E[{D_{{\rm{prop\_\rmnum{2}}}}}].
\end{aligned}
\end{equation}

To complete the proof, substitute the Eq. \eqref{dprop1}, \eqref{Tpause}, \eqref{Dpause} and \eqref{Thoping} into Eq. \eqref{Tfderi} and \eqref{Dfderi}.
\end{IEEEproof}

\begin{theorem}
\label{ipsvmimotheorem}
The IPS achieved by the proposed virtual MIMO approach is:
\begin{equation}
\label{vpmimo}
{V_{{\rm{prop\_VMIMO}}}} = \frac{{\left( {\frac{{{\sigma _2}}}{{{\sigma _1}}} + {\sigma _1}} \right) \times \frac{{\frac{1}{{\frac{1}{{\lambda {r^2}}} + \frac{1}{R}}}}}{{{p_b}}} + \frac{{{\sigma _2}}}{{{\sigma _1}}} \cdot \frac{1}{\lambda }}}{{\left( {\frac{{{\sigma _2}}}{{{\sigma _1}}} + {\sigma _1}} \right) \times \frac{\tau }{{{p_b}}} + \frac{1}{{{\sigma _1}}} \cdot \frac{1}{{2v\lambda }}}},
\end{equation}
where $\sigma _1$ and $\sigma _2$ are given by Eq. \eqref{s1} and Eq. \eqref{s2} respectively. $p_b$ is given by Eq. \eqref{pbequa}.
\end{theorem}
\begin{IEEEproof}
The main steps of the proof are listed in Table \ref{IPStable} for explicitness and brevity.
\begin{table}[!h]
\caption{Overview of the proof}\label{IPStable}
  \renewcommand{\arraystretch}{1.5}
  \centering
  \begin{tabular}{|c|c|c|}
  \hline
  \multicolumn{3}{|c|}{IPS $V_{{\rm{prop\_VMIMO}}}: Eq. \eqref{vp}$} \\
  \hline
  Distance $E[D]{:}$             & \multicolumn{2}{c|}{Time $E[T]$}  \\ \cline{2-3}
  $ $    &Waiting time $E[T_{\rm{prop}}]{:}$   &  $E[T_{\rm{stop}}]{:}$   \\
  $Eq. \eqref{Dfeq}$         & $Eq. \eqref{Tfeq}$&  $Eq. \eqref{Tstopeq}$  \\ \hline
  \end{tabular}
\end{table}
\end{IEEEproof}

\section{Asymptotic Analysis and Key Observations}
\label{furtheranalysis}
Based on the closed-form IPS given by Eq. \eqref{vpmimo}, in the following, we will conduct further asymptotic analysis to reveal the impacting factors on IPS, such as network density and moving speed, etc., so as to get more insight and clearer picture on highway information propagation.

\subsection{Impact of the Traffic Density}
\subsubsection{Low Density Regime}
In the low density regime, we have the following observations.

\begin{corollary}
\label{corolow}
When the traffic density is low, the IPS increases cubically with respect to the total vehicle density. Extremely, as $\lambda \to 0$, $V_{\rm{prop\_VMIMO}} \to v$.
\end{corollary}
\begin{IEEEproof}
According to \emph{Lemma} \ref{pb}, $p_b^{-1}$ increases quasi-exponentially with $\lambda$, which makes it a dominant term influencing $V_{\rm{prop\_VMIMO}}$. Therefore, by keeping the dominant terms, we can approximate $V_{\rm{prop\_VMIMO}}$ to:
\begin{equation}
\label{vpomit}
\begin{aligned}
&V_{\rm{prop\_VMIMO}}\approx  \\
&\frac{{(\lambda {\lambda _f} + {\lambda _r}^2)\frac{{\lambda {r^2}R}}{{\lambda {r^2} + R}}p_b^{ - 1} + ({\lambda _f} - {\lambda _r}){\lambda _r}\frac{{\lambda {r^2}R}}{{\lambda {r^2} + R}} + {\lambda _f}}}{{(\lambda {\lambda _f} + {\lambda _r}^2)\tau p_b^{ - 1} + ({\lambda _f} - {\lambda _r}){\lambda _r}\tau  + \frac{\lambda }{{2v}}}}.
\end{aligned}
\end{equation}
Without loss of generality, we set $\lambda_r=\lambda_f = \frac{\lambda}{2}$. With some re-organization, we have:
 \begin{equation}
 \label{lreqlf}
 V_{\rm{prop\_VMIMO}}\approx\frac{{3\lambda p_b^{ - 1}\frac{{\lambda {r^2}R}}{{\lambda {r^2} + R}} + 2}}{{3\lambda \tau p_b^{ - 1} + \frac{2}{v}}}.
 \end{equation}
 Then, based on the result of \emph{Lemma} \ref{pb}, Eq. \eqref{lreqlf} becomes:
\begin{equation}\label{vapprox}
V_{\rm{prop\_VMIMO}}  \approx \frac{3\lambda \Theta \left( {{e^{\lambda \beta }}} \right)\frac{{\lambda {r^2}R}}{\lambda {r^2} + R}+2}{{3\lambda \tau \Theta \left( {{e^{\lambda \beta }}} \right) + \frac{2}{v}}}.
\end{equation}
When $\lambda$ is sufficiently small, Eq. \eqref{vapprox} can be further approximated by:
\begin{equation}
\label{lowdensityapprox}
\begin{aligned}
V_{\rm{prop\_VMIMO}}&\approx\frac{3\lambda \left(\omega \lambda \beta\right){\lambda {r^2}}+2}{3\lambda \left(\omega \lambda \beta\right)\tau+\frac{2}{v}} \\
&\approx\frac{3\omega \beta r^2 v}{2} \lambda^3+v,
\end{aligned}
\end{equation}
where the first approximation follows from: $\frac{{\lambda {r^2}R}}{{\lambda {r^2} + R}} = \frac{1}{\frac{1}{\lambda r^2}+\frac{1}{R}}\approx \lambda r^2$, and $\Theta \left( {{e^{\lambda \beta }}} \right) \approx \omega \lambda \beta$, where $\omega$ is a constant, meanwhile the second approximation holds by omitting the first term in the denominator as $\frac{2}{v}$ is the dominant term, in other words, there exist a sufficient small $\lambda_0$ which satisfies that when $\lambda<\lambda_0$, $\frac{2}{v} \gg 3 \tau \omega \beta \lambda^2 $.  Extremely, as $\lambda \to 0$, obviously, there is $V_{\rm{prop\_VMIMO}} \to v$.
\end{IEEEproof}

\emph{Remarks:} In a word, the cubical growth of IPS is caused by the positive feedback loop as follows: more senders lead to longer transmission range, which in turn includes more senders as contributors. Specifically, in the very low vehicle density (sparse) case, the IPS is determined by vehicle speed $v$ because there is little packet forwarding. Intuitively, as the vehicle density starts to grow, the cubical growth of IPS comes from the following interrelating facts: 1) the average number of vehicles that fall within the detection range R increases as the vehicle density grows, 2) the resultant combined SNR grows approximately linearly in the number of informed vehicles that form the virtual MIMO, and 3) as the combined SNR increases, the effective propagation range (i.e., distance propagated during one time slot) increases accordingly, which in turn makes more vehicles informed within the detection range $R$ and thus increases the opportunities to form new virtual MIMO. This way, the information propagation is sped up cubically.

\subsubsection{High Density Regime}
Correspondingly, in the high density regime, we have the following observation.
\begin{corollary}
\label{corohigh}
When the traffic density is high, there is
\begin{equation}
\label{vapproxhigh}
V_{\rm{prop\_VMIMO}} \approx \frac{1}{\tau }\frac{{\lambda {r^2}R}}{{\lambda {r^2} {+} R}}.
\end{equation}
Extremely, as $\lambda \to \infty$, $V_{\rm{prop\_VMIMO}} \to \frac{R}{\tau}$.
\end{corollary}
\begin{IEEEproof}
First, follow the same approximation line as in the low density case, which also leads to Eq. \eqref{vapprox}.  As $\lambda$ is sufficiently large, we can safely omit the constant terms, and yield Eq. \eqref{vapproxhigh}. Then, $\frac{R}{\tau}$ directly follows as $\lambda \to \infty$.
\end{IEEEproof}

\emph{Remarks:} As the vehicle density further increases, the forwarding process starts to dominate, i.e., the propagation is seldomly interrupted by large gaps. However, the packet can propagate no further than the detection range per time slot, since it cannot be detected by any vehicle beyond $R$. Therefore, the IPS will keep increasing until it reaches the maximum value $\frac{R}{\tau}$, which also implies that increased R, i.e., a stronger detection ability can also boost the information propagation by exploiting more power gain, exhibiting a slower converging process (by checking the derivative with respect to $\lambda$) as well as a higher upper bound.
\subsection{Impact of the Vehicle Speed}
Apart from the vehicle density, we also examine the impact of mobility on IPS.
\begin{corollary}
\label{vv}
The IPS grows with the vehicle speed. Extremely, we have:
\begin{equation}
v \to 0 \Longrightarrow V_{\rm{prop\_VMIMO}} \to 0
\end{equation}
\begin{equation}
v \to \infty \Longrightarrow V_{\rm{prop\_VMIMO}} \to \frac{1}{\tau}\frac{{\lambda {r^2}R}}{{\lambda {r^2} {+} R}} + \frac{\frac{{{\sigma _2}}}{{{\sigma _1}}} \cdot \frac{1}{\lambda }}{\left( {\frac{{{\sigma _2}}}{{{\sigma _1}}} + {\sigma _1}} \right) \cdot \frac{\tau }{p_b}}.
\end{equation}
\end{corollary}
\begin{IEEEproof}
 As $v \to 0$, $\frac{1}{2v\lambda \sigma_1}$ in Eq. \eqref{vpmimo} tends to infinity, straightforwardly, we have $V_{\rm{prop\_VMIMO}} \to 0$. In contrast, as $v \to +\infty$, $\frac{1}{2v\lambda \sigma_1} \to 0$ and thus can be omitted, which directly leads to the result in \emph{Corollary} \ref{vv}.
\end{IEEEproof}

\emph{Remarks:} Intuitively, in the low mobility case, once the beacon is blocked, the propagation process is terminated. Thereby the average IPS in the long-run can be viewed as zero. As $v$ increases, more dynamics are introduced into the network topology, thus more communication opportunities are created to accelerate the information propagation. However, the performance is still constrained by the network density, with the limit indicating fastest possible spreading case where the latency for restoring connectivity is negligible due to infinite $v$.

\subsection{Comparison with the Conventional Flooding}
In this subsection, we show the advantage of our proposed scheme by theoretically comparing it with the conventional flooding scheme, which no longer adopts the signal-combining mechanism as described in Section \ref{vmimo}. In other words, at each time slot, only those who are inside the transmission range $r$ of an informed vehicle can successfully decode the packet.

We may follow the same line of evaluating the IPS of the virtual MIMO scheme. The differences are that, the expected distance covered in hop state is replaced by $r$, and the probability of being blocked equals $e^{-\lambda r}$, which straightforwardly leads us to:

\label{IPSconvsec}
\begin{theorem}
\label{ipsconvtheo}
The IPS achieved by the conventional flooding Scheme is:
\begin{equation}
\label{vpconveq}
{V_{{\rm{prop\_conv}}}} = \frac{{\left( {\frac{{{\sigma _2}^{'}}}{{{\sigma _1}^{'}}} + {\sigma _1}^{'}} \right){e^{\lambda r}}r + \frac{{{\sigma _2}^{'}}}{{{\sigma _1}^{'}}} \cdot \frac{1}{\lambda }}}{{\left( {\frac{{{\sigma _2}^{'}}}{{{\sigma _1}^{'}}} + {\sigma _1}^{'}} \right){e^{\lambda r}} \tau + \frac{1}{{{\sigma _1}^{'}}} \cdot \frac{1}{{2v\lambda }}}},
\end{equation}
where ${\sigma _1^{'}}{\rm{ = }}\frac{{{\lambda _r}}}{{{\lambda _r}{\rm{ + }}{\lambda _f}}}{\rm{ - }}\frac{{{\lambda _r}}}{{{\lambda _r}{\rm{ + }}{\lambda _f}}}{e^{-\lambda r}}$, ${\sigma _2^{'}}{\rm{ = }}\frac{{{\lambda _f}}}{{{\lambda _r}{\rm{ + }}{\lambda _f}}}{\rm{ + }}\frac{{{\lambda _r}}}{{{\lambda _r}{\rm{ + }}{\lambda _f}}}{e^{-\lambda r}}$.
\end{theorem}

We denote by $g$ the IPS gain, defined as the the ratio between $V_{\rm{prop\_VMIMO}}$ and $V_{\rm{prop\_conv}}$.
\begin{corollary}
\label{coroipsgain}
In the high density regime, $g \approx \frac{\lambda r R}{\lambda r^2 +R}$. Extremely, we have:
\begin{equation}
\lambda \to +\infty, g \longrightarrow \frac{R}{r}.
\end{equation}
\end{corollary}
\begin{IEEEproof}
Thanks to the symmetry between the Eq. \eqref{vpconveq} and \eqref{vpmimo}, we can similarly derive:

As $\lambda \to +\infty$, $V_{\rm{prop\_conv}} \to \frac{r}{\tau}$. Hence, with the result in \emph{Corollary} \ref{corohigh}, simply taking the ratio of these two yields \emph{Corollary} \ref{coroipsgain}.
\end{IEEEproof}

\emph{Remarks:} Surprisingly, we find that the performance gain under extreme dense case is determined by the ratio of detection range $R$, and the transmission range $r$, which further suggests the potential benefits of a stronger detection ability.                                                                                                                                                                                                                                       \section{Simulation Results}
\label{simulation}
In this section, intensive information propagation experiments are conducted to verify the correctness of the derived theoretical results. The simulations follow precisely the model described in Section \ref{systemmodel}. We generate two opposing traffic which are Poisson distributed with density $\lambda_f$ and $\lambda_r$ vehicles/m and measure the IPS by selecting a sufficient remote source-destination pair, taking the ratio of the propagation distance over the corresponding delay, and averaging over multiple iterations of randomly generated traffic. The transmission range $r$ and the detection range $R$ are set to 200m and 600m by default and $\tau$ is set to 25ms.
\begin{figure}[!ht]
\centering
\includegraphics[angle=0,scale=0.35]{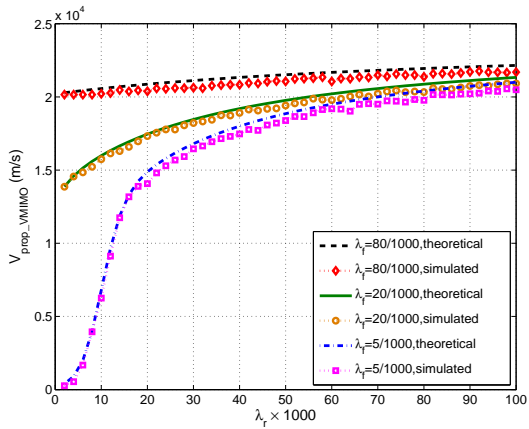}
\caption{IPS with respect to vehicle density.}
\label{vpdensitydifferl2}
\end{figure}

\begin{figure}[!ht]
\centering
\includegraphics[angle=0,scale=0.35]{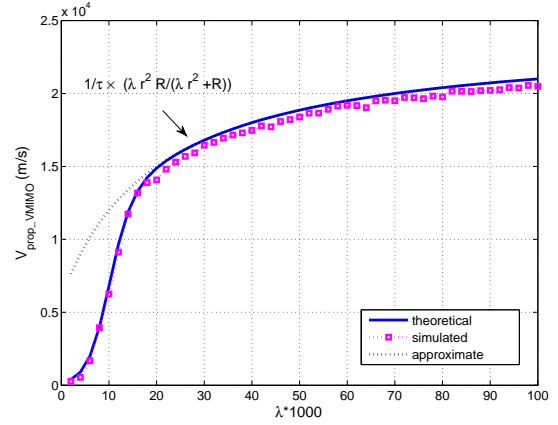}
\caption{Approximation of IPS in the high density regime.}
\label{approxhighlambda}
\end{figure}
In Fig. \ref{vpdensitydifferl2}, we vary $\lambda_r$, and plot the resulting IPS by the proposed virtual MIMO approach under several $\lambda_f$ settings. A close match between theoretical and simulation results can be observed, which confirms the analysis of IPS presented in Section \ref{IPSvmimosec}. Fig. \ref{approxhighlambda} shows that the IPS increases with the total vehicle density, yet the curve tends to flatten out, consisting with the predicted approximation of $\frac{1}{\tau }\frac{{\lambda {r^2}R}}{{\lambda {r^2} + R}}$, and finally approach $\frac{R}{\tau}$, as expected in \emph{Corollary} \ref{corohigh}. It is also observed that, during the primary stage, particularly in the case of a small $\lambda_r$, the sharp increase of the IPS consists with a ``$y=a \lambda^3+b$'' curve pattern, which confirms the cubical growth in \emph{Corollary} \ref{corolow}.

\begin{figure}[!ht]
\centering
\includegraphics[angle=0,scale=0.35]{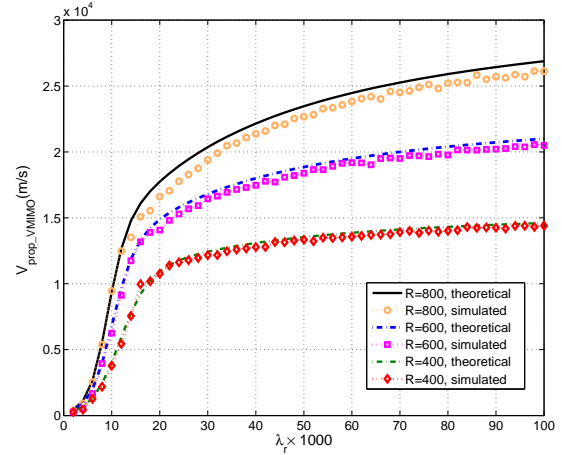}
\caption{IPS under different $R$'s.}
\label{changeR}
\end{figure}

Furthermore, Fig. \ref{changeR} demonstrates the impact of the detection range $R$ on the achieved IPS. As expected, larger $R$ results in faster IPS increase as well as higher convergence value, indicating a great benefit of stronger detecting ability.

\begin{figure}[!ht]
\centering
\includegraphics[angle=0,scale=0.35]{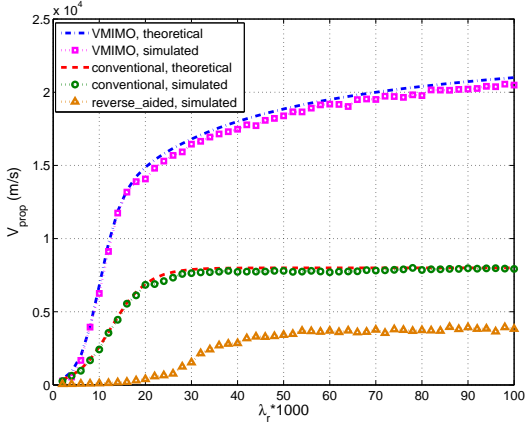}
\caption{IPS Comparisons with respect to vehicle density.}
\label{complambda}
\end{figure}

\begin{figure}[!ht]
\centering
\includegraphics[angle=0,scale=0.35]{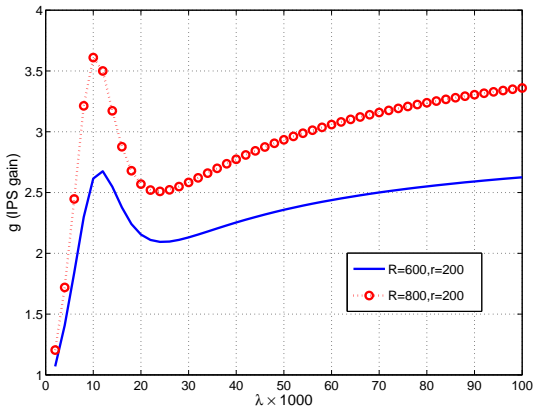}
\caption{IPS gain over conventional flooding scheme with vehicle density.}
\label{ipsgain}
\end{figure}

\begin{figure}[!ht]
\centering
\includegraphics[angle=0,scale=0.35]{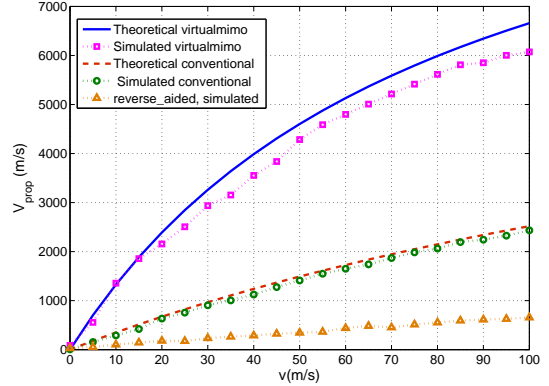}
\caption{IPS Comparisons with respect to vehicle speed.}
\label{vpropv}
\end{figure}

In Fig. \ref{complambda} and Fig. \ref{vpropv}, we compare the achieved IPS of our virtual-MIMO-enabled scheme with other baseline methods, in terms of vehicle density and vehicle speed, respectively. Other than the conventional flooding scheme, a unicast-based routing scheme proposed by \cite{exactphasetransi}, referred to as ``reverse\_aided'' approach, is also conducted here for comparison purpose. Unlike the conventional flooding, the ``reverse\_aided'' will keep the beacon in the head until a sufficiently large cluster is encountered to bridge the gap. In both experiments, our virtual MIMO approach exhibits a remarkable IPS gain over others. For instance, when the traffic density is around 0.003 vehicles/m, the ``reverse\_aided'' approach can only achieve an IPS of about 2000m/s, and the conventional flooding performs better with an IPS of 8000m/s, but already reaches its limit, while our virtual MIMO approach yields 16000m/s, and still has room for uprise; The IPS gain is further plotted semi-logarithmically in Fig. \ref{ipsgain}, where a sharp increase, which nearly coincides with a straight dotted line, can be observed as the vehicle densities increases from zero, and the curve tends to flatten out as the traffic becomes extremely denser, approaching $\frac{R}{r}$, which confirms \emph{Corollary} \ref{coroipsgain}. Moreover, Fig. \ref{ipsgain} also implies the existence of an optimal $\lambda$, which unfortunately, is difficult to be derived in closed-form, but can be numerically obtained.

Again, the proposed scheme is superior when varying the vehicle speed (e.g. a huge IPS enhancement from 1500m/s by conventional approach (or worse, 500m/s by ``reverse\_aided'' approach) to about 4500m/s  by the virtual MIMO approach). Moreover, notice that Fig. \ref{vpropv} confirms remarks on \emph{Corollary} \ref{vv}, indicating that increased mobility accelerates information propagation, though not unboundedly.

\section{Conclusions and Future Work}
\label{conclusion}
In this paper, we study the information propagation speed in the bidirectional highway scenario. We proposed a novel virtual-MIMO-enabled information dissemination scheme, in which the vehicles opportunistically form virtual antenna arrays to simultaneously transmit signals while others independently and aggressively combine as many signals as they can detect, to hopefully decode the emergency message at a further distance. It is shown that the scheme effectively boosts one-hop transmission range and exhibits significant IPS gain over others. Both closed-form IPS results of the proposed scheme and the conventional flooding scheme are derived. Although our current work is still theoretical-based, and the model is still too simplified for practical implementation, our theoretical results revealed some insightful properties
of the derived IPS, which may help designing and evaluating appropriate routing protocols in VANET. It is implied that increased vehicle density and network mobility will both aid in information propagation. In our future work, we intend to design some MAC layer protocols and physical layer techniques which perfectly match our proposed scheme, to create a more completed and implementable cross-layer framework. Further, we hope to extend the work to a two-dimensional scenario.

\appendix
\subsection{Proof of Lemma \ref{pb}} \label{proof_lemma1}

Assume that the nearest uninformed vehicle \textbf{u} (located at $x_u$) is able to detect a random number of $N$ transmitting vehicles. Denote $\textrm{B}_1$ the event that the propagation is blocked, $\textrm{B}_2$ the event that there is no informed vehicle within $r$ distance of \textbf{u}. Applying law of total probability, we write $p_b$ as follows:
\begin{equation}
\begin{aligned}
p_b&=\textrm{Pr}\{\textrm{B}_1|\textrm{B}_2\}\cdot \textrm{Pr}\{\textrm{B}_2\}+\textrm{Pr}\{\textrm{B}_1|\overline{\textrm{B}_2}\}\cdot \textrm{Pr}\{\overline{\textrm{B}_2}\} \\
   &=\textrm{Pr}\{\textrm{B}_1|\textrm{B}_2\}\cdot e^{-\lambda r}.
\end{aligned}
\end{equation}

Therefore, our remaining effort is to quantify the block probability under the condition that $N$ informed vehicles are within range $\left(x_u-R, x_u-r\right)$ where $N$ is Poisson distributed with density $\lambda_N = \lambda \left(R-r\right)$ (note that $\lambda$ is the number of vehicles per unit length). The received SNR at \textbf{u} is
\begin{equation}
\gamma=\frac{\alpha P_t}{N_0}\sum \limits_{i=1}^{N} \frac{1}{d_i^2},
\end{equation}
where $d_i$ ($i=1,2,...,N$) denotes the corresponding distance between each transmitter and \textbf{u}. Due to the Poisson distribution property, for adequately large $R$, $d_i$ is approximately uniformly distributed within $(r, R)$, thus we have $E[\frac{1}{d_i^2}]=\frac{1}{rR}$ and the second moment $E[\left(\frac{1}{d_i^2}\right)^2]=\frac{\lambda\left(R^3-r^3\right)}{3\left(R-r\right) R^3 r^3}$.

Now applying the following result (\cite{asymnormal}, page 1160, Example \rmnum{1}): if an RV $Y$ has the form $Y=\sum\limits_{i=1}^{N}X_i$, where $N$ is Poisson distributed with density $\lambda$ and $X_i$ ($i=1,2,...,N$) are i.i.d. RVs with mean $E[X]=a$ and second moment $E[X^2]=b^2$, then $Y$ is asymptotically normal with mean $\lambda a$ and deviation $b \sqrt{\lambda}$, we can conclude that $\gamma$ is approximately Gaussian distributed with mean $\frac{\alpha P_t}{N_0}\lambda_N E[\frac{1}{d_i^2}]$ and deviation $\frac{\alpha P_t}{N_0}\sqrt{\lambda_N E[\left(\frac{1}{d_i^2}\right)^2]}$. Hence, the conditional block probability can be directly derived from the distribution of $\gamma$, which leads to the first equation of Eq.(\ref{pbequa}) after some manipulation. Given the first equation holds, the second equation of Eq.(\ref{pbequa}), i.e., $p_b = \Theta \left(e^{-\beta \lambda}\right)$, is obvious.


\subsection{Lemma \ref{Dhopping} and Its Proof} \label{proof_lemma5}
\begin{lemma}
\label{Dhopping}
For the proposed virtual MIMO-based scheme, the expected distance that the beacon can maximally propagate in one time slot, is given by:
\begin{equation}
\label{Dhopeq}
E\left[D_{\rm{mprop}}\right]=\frac{1}{\frac{1}{\lambda r^2}+\frac{1}{R}}.
\end{equation}
\end{lemma}
\begin{IEEEproof}
First, we consider a general scene where the distance between the head (located at $x_0$) and the nearest uninformed vehicle ahead, referred to as `receiver', is $d$, and the receiver is able to detect $n$ vehicles at the time. Denote $d_i (i=1,2,3...,n)$ the distance between each detected vehicle and the receiver. Recalling the conditional distribution of the arrival times proposition in \cite{probabilitymodels} (Page 336, Proposition 5.4), we argue that, $d_i\left(i{=}1,2,...,n-1\right)$ are identically independently distributed over the interval $\left(x_0-\left(R-d\right), x_0\right)$, which directly gives the probability density function (pdf) of $z_i=\frac{1}{d_i^2}$: $f_{Z_i}\left(z\right)=\frac{1}{2\left(R-d\right)}z^{{-}\frac{3}{2}}$. By taking the approximation $\sum\limits_{i=1}^{n-1} {\frac{1}{d_i^2}}\approx \left(n-1\right)E\left[Z_i\right]$, we can approximate the received power as:
\begin{equation}
\label{rxpower}
P_r \approx {\alpha P_t} \left(\frac{1}{d^2} + \frac{n-1}{dR}\right).
\end{equation}

Ideally, we assume that the beacon can maximally reach a point $D_m$-distance away. Based on Eq. \eqref{rxpower}, the received power $P_{rm}$ clearly satisfies:
\begin{equation}
\begin{aligned}
P_{rm}=N_0 \gamma_{dec}&=\frac{\alpha P_t}{r^2} \\
                       &=\alpha P_t \left(\frac{1}{D_m^2}+\frac{N_m-1}{D_m R}\right) \\
                       &\approx \alpha P_t \frac{N_m}{D_m R},
\end{aligned}
\end{equation}
which gives:
\begin{equation}
D_m \approx \frac{N_m r^2}{R}.
\end{equation}
Taking the expectation of both sides with $E\left[N_m\right]=\lambda \left(R- E\left[D_m\right]\right)$ further yields (with some necessary rearrangement):
\begin{equation}
\label{D_m}
E\left[D_m \right] \approx \frac{\lambda R r^2} {R+\lambda r^2}
\end{equation}
Note that the actual distance $D_h$ traversed by the beacon cannot exceed $D_m$. In fact, we have:
\begin{equation}
D_h = \sum \limits_{i=1}^{N_m} c_i | N_m=\max \left(m: c_1+c_2+\cdot\cdot\cdot+c_m \le D_m\right),
\end{equation}
where $c_i (i=1,2,3...)$ denote the i.i.d exponential inter-vehicle distances. Borrowing results from \cite{probabilitymodels} (Chapter 7, page 456), we argue that:
\begin{equation}
\begin{aligned}
E\left[D_h\right] &= E\left[\sum \limits_{i=1}^{N_m} C_i\right] = E\left[E\left[\sum \limits_{i=1}^{N_m} C_i|N_m\right]\right] \\
                  &= E\left[N_m E\left[C\right]\right] = E\left[N_m\right] E\left[C\right] \\
                  &= \lambda E\left[D_m\right] \cdot \frac{1}{\lambda} = E\left[D_m\right],
\end{aligned}
\end{equation}
where the fifth equality holds due to the exponentially-distributed inter-vehicle distance. Thus, $E\left[D_h\right]$ is obtained directly form Eq. \eqref{D_m}.
\end{IEEEproof}

\bibliographystyle{IEEEtran}

\begin{thebibliography}{1}
\bibitem{model1}
T.D.C.~Little, and A.~Agarwal, ``An Information Propagation Scheme for Vehicular Networks," in \emph{Proc. IEEE ITSC}, pp. 155-160, 2005.

\bibitem{model2}
T.~Nadeem, P.~Shankar, and L.~Iftode, ``A Comparative Study of Data Dissemination Models for VANETs," in \emph{Third Annual International Conference on Mobile and Ubiquitous Systems£º Networking Services}, pp. 1-10, 2006.

\bibitem{model3}
W.~Leutzbach, ``Introduction to the Theory of Traffic Flow," \emph{Springer-Verlag}, 1988.

\bibitem{add1}
J.~Yang, and Z.~Fei, ``Broadcasting with Prediction and Selective Forwarding in Vehicular Networks," in \emph{International Journal of Distributed Sensor Networks}, 2013.

\bibitem{add2}
J.~Yang, and Z.~Fei, ``Statistical Filtering Based Broadcast Protocol for Vehicular Networks," in \emph{Proc. IEEE ICCCN}, pp. 1-6, 2011.

\bibitem{poisson}
F.~Bai, and B.~Krishnamachari, ``Spatio-temporal variations of vehicle traffic in VANETs: facts and implications," in \emph{Proc. ACM international workshop on Vehicular Internetworking}, 2009.

\bibitem{DSRC}
``Dedicated Short Range Communications (DSRC) Home," [Online]. http://www.leearmstrong.com/DSRC/DSRCHomeset.htm

\bibitem{IEEE802}
D.~Jiang, and L. Delgrossi, ``IEEE802.11p: Towards an international standard for wireless access in vehicular environments,'' in \emph{IEEE VTC}, 2008.

\bibitem{HZ1}
H.~Zhang, Z.~Zhang, and H.~Dai, ``Gossip-based Information Spreading in Mobile Networks," \emph{IEEE Transactions on Wireless Communications}, vol. 12, no.11, pp. 5918-5928, Nov. 2013.

\bibitem{HZ2}
H.~Zhang, Y.~Huang, Z.~Zhang, and H.~Dai, ``Mobile Conductance in Sparse Networks and Mobility-Connectivity Tradeoff," in Proc. IEEE International Symposium on Information Theory (ISIT), pp. 256-260, 2014.

\bibitem{model4}
N.~Wisitpingphan, F.~Bai, P.~Mudalige, and O.~Tonguz, ``On the Routing Problem in Disconnected Vehicular Ad-hoc Networks," in \emph{Proc. IEEE INFOCOM}, 2007.

\bibitem{model5}
G.~Yan, N.~Mitton, and X.~Li, ``Reliable routing in vehicular ad hoc networks," in \emph{The 7th International Workshop on Wireless Ad hoc and Sensor Networking}, Genoa, Italie, 2010.

\bibitem{congestioncontrol}
S.~Djahel, and Y.~Ghamri-Doudane, ``A Robust Congestion Control Scheme for Fast and Reliable Dissemination of Safety Messages in VANETs," in \emph{Proc. IEEE WCNC}, pp. 2264-2269, 2012.

\bibitem{tlc}
Y.~Zhuang, J.~Pan, Y.~Luo, and L.~Cai, ``Time and Location-critical Emergency Message dissemination for Vehicular Ad-hoc Network," in \emph{IEEE J.Sel. Areas Commun.}, vol. 29, no. 1, pp. 187-196, 2011.

\bibitem{upblb1}
A.~Agarwal, D.~Starobinski, and T.D.~Little, ``Analytical model for message propagation in delay tolerant vehicular ad hoc networks," in \emph{Proc. IEEE VTC}, pp. 3067-3071, 2008.

\bibitem{upblb2}
A.~Agarwal, and T.D.~Little, ``Impact of asymmetric traffic densities on delay tolerant vehicular ad hoc networks,¡± in \emph{Proc. IEEE VNC}, pp. 1-8, 2009.

\bibitem{exactphasetransi}
E.~Baccelli, P.~Jacquet, B.~Mans, and G.~Rodolakis, ``Information propagation speed in bidirectional vehicular delay tolerant networks," in \emph{Proc. IEEE INFOCOM}, pp. 436-440, 2011.

\bibitem{varyradiorange}
E.~Baccelli, and P.~Jacquety, ``Multi-lane Vehicle-to-Vehicle Networks with Time-Varying Radio Ranges: Information Propagation Speed Propertie," in \emph{Proc. IEEE ISIT}, 2013.

\bibitem{wu}
H.~Wu, R.M.~Fujimoto, G.F.~Riley, and M.~Hunter, ``Spatial propagation of information in vehicular networks," \emph{IEEE Trans. on Vehicular Technology}, vol. 58, no. 1, pp. 420-431, 2009.

\bibitem{varyv1}
Z.~Zhang, G.~Mao, and B.D.~Anderson, ``On the information propagation speed in mobile vehicular ad hoc networks," in \emph{Proc. IEEE GLOBECOM}, pp.1-5, 2010.

\bibitem{varyv2}
Z.~Zhang, G.~Mao, and B.D.~Anderson, ``On the Information Propagation Process in Multi-lane Vehicular Ad-hoc Networks," in \emph{Proc. IEEE ICC}, 2012.

\bibitem{faster}
H.~Wu, Z.~Zhang, and H.~Zhang, ``Faster Information Propagation on Highways: a virtual MIMO approach," \emph{Proc. IEEE GLOBECOM}, pp.1654 - 1660, 2014.


\bibitem{sync1}
F.~Sivrikaya, and B.~Yener, ``Time synchronization in sensor networks: A Survey," \emph{IEEE Network}, Vol. 18, No. 4, pp. 45-50, 2004.


\bibitem{sync2}
F.~Guo, D.~Li, H.~Yang, and L.~Cai,``A novel timing synchronization method for distributed MIMO-OFDM system,'' in \emph{Proc. IEEE VTC} vol. 4, pp. 1933-1936, 2006.

\bibitem{sync3}
Y.~Cheng, Y.~Jiang, and X.-H. You,``Preamble design and synchronization algorithm for cooperative realy systems,'' in \emph{Proc. IEEE VTC}, 2009.

\bibitem{wc}
A. Goldsmith, ``Wireless Communications,'' Cambridge University Press, 2005.

\bibitem{asymnormal}
H.~Robbins, ``The Asymptotic Distribution of the Sum of a Random Number of Random Variables," \emph{Bulletin of the American Mathematical Society}, No. 54, pp. 1151-1161, 1948.

\bibitem{probabilitymodels}
S.M.~Ross, \emph{Introduction to Probability Models}, Tenth Edition, Elsevier Pte Ltd, 2010.
\end{thebibliography}

\end{document}